\documentclass[twoside]{article}
\usepackage{fleqn,espcrc2,epsfig}

    \newcommand{\scr}{\rm\scriptscriptstyle}

\newcommand{\AmS}{{\protect\the\textfont2
  A\kern-.1667em\lower.5ex\hbox{M}\kern-.125emS}}

\title{Quantum algebras in phenomenological description
                        of particle properties}

\author{A.M. Gavrilik\address[MCSD]{Bogolyubov
        Institute for Theoretical Physics, \\
        03143, Kiev, Ukraine}
        \thanks{
                This research was partly supported by the U.S.
                Civilian Research and Development Foundation
                Grant UP1-2115.
                      }}

\begin{document}

\begin{abstract}
Quantum and $q$-deformed algebras find their application not only in
mathematical physics and field theoretical context, but also in
phenomenology of particle properties.  We describe  (i) the use of
quantum algebras $U_q(su_n)$ corresponding to Lie algebras of
the groups $SU_n$, taken for flavor symmetries of hadrons,
in deriving new high-accuracy hadron mass sum rules, and  (ii) the
use of (multimode) $q$-oscillator algebras along with $q$-Bose gas
picture in modelling the properties of the intercept $\lambda$ of
two-pion (two-kaon) correlations in heavy-ion collisions,
as $\lambda$ shows sizable observed deviation from the expected
Bose-Einstein type behavior. The deformation parameter $q$ is
in case (i) argued and in case (ii) conjectured to be connected
with the Cabibbo angle $\theta_{\scr{C}}$.

\vspace{1pc}
\end{abstract}

\maketitle


\section{Introduction}

Quantum groups and quantum or $q$-deformed
algebras   \cite{dr-j,frt}, whose basic mathematical aspects and
diverse quantum physical applications are intensively studied
for about a decade and half, will belong to most important
and perspective tools of research in the 3rd millenium, too.

In this talk, meant as a mini-review, we concentrate on two examples
of applying $q$-algebras to phenomenology of hadrons.  Within the
first one initiated in   \cite{ga-ser} and developed in subsequent
papers, the $q$-analogs $U_q(su_n)$ of the Lie algebras of groups
$SU_n$ are adopted for hadronic flavor symmetries in order to derive
new results concerning hadron masses and mass sum rules. Basic tool of
this aproach is the representation theory of the $q$-algebras
$U_q(su_n)$  \cite{dr-j,frt}, and in sections 2-8 we discuss a number
of results, including unexpected implications: a possibility (gained due
to use of $q$-algebras) to label different flavors topologically -
by knot invariants;  a direct link of deformation parameter
to the Cabibbo angle   \cite{cabi}, etc.
   In the second part of the talk (section 9) we consider a
usage of the algebras of $q$-deformed oscillators,
within a model of $q$-Bose gas, for effective
description of unusual (non-Bose type) behavior of
two-particle correlations of hadrons (pions or kaons) produced
and registered in heavy ion collisions.

\section{Vector mesons: $q$-deformation vs mixing}

We use (see    \cite{ga-ser,ga-bre,ga-uzh})  Gelfand-Tsetlin basis
vectors for meson states from $(n^2\!-\!1)$-plet of '$n$-flavor'
$U_q(u_n)$ embedded into $\{(n\!+\!1)^2\!-\!1\}$-plet of
'dynamical' $U_q(u_{n+1});\ $ construct mass operator $\hat {M_n}$,
invariant under the 'isospin+hypercharge' $q$-algebra $U_q(u_2)$,
from the generators of dynamical algebra $U_q(u_{n+1})$
(e.g., $\hat {M_3} = M_0{\bf 1}+{\gamma}_3 A_{34}A_{43}
+{\delta}_3 A_{43}A_{34}$);  calculate expressions for masses
$m_{V_i}\equiv\langle V_i|\hat{M_3}|V_i\rangle $ -
these involve $M_0$, the parameters $\gamma_3, {\delta}_3$,
and the $q$-parameter.
In particular, for $n=3$ one obtains
\vspace{-1.0mm}
\begin{eqnarray}
{} & m_{\rho}=M_0,  \hspace{4mm}  m_{K^*}=M_0 -{\gamma}_3,
                                        \nonumber \\
{} & \hspace{-6mm}
m_{{\omega}_8}= M_0 - 2 {[2]_q}/{[3]_q} {\gamma}_3,     \label{(1)}
\end{eqnarray}
where $[x]_q \equiv[x]= \frac{q^x - q^{-x}}{q - q^{-1}}$ is
the $q$-number 'deforming' a number $x$ and, to have equal masses
for particles/antiparticles, $\delta_3\!=\!\gamma_3$ was set.
$q$-Dependence appears only in the mass of ${\omega }_8$ (isosinglet
in $U_q(su_3)$-octet).  Excluding $M_0, \gamma_3$, the $q$-analog of
Gell--Mann - Okubo (GMO) relation is      \cite{ga-ser} :
\vspace{-1.0mm}
\begin{equation}
[3]_q m_{{\omega }_8} + (2[2]_q-[3]_q) m_{\rho }
                    =  2[2]_q m_{K^*}\ .               \label{(2)}
\end{equation}
In the limit $q\!=\!1$ (then, $\frac{[3]_q}{[2]_q}=\frac32$),
this reduces to
usual GMO formula $3m_{{\omega}_8}\!+\! m_{\rho} = 4 m_{K^*}$
which needs singlet mixing                                \cite{8-fold}.
However, it also yields
\begin{equation} m_{{\omega}_8} + m_{\rho} = 2 m_{K^*}
               \hspace{8mm}  {\rm if}  \hspace{3mm}   q=e^{i\pi /5}                                   \label{(3)}
\end{equation}
(then,  $[3]_q=[2]_q$).
With $m_{{\omega}_8}\equiv m_{\phi}$, and no mixing, eq.(3) coincides
with nonet mass formula of Okubo  \cite{oku-fi} agreeing {\it ideally}
with data  \cite{PDG}.
The deformation angle $\frac{\pi}{5}$, see (3), coincides remarkably
with $\omega$-$\phi$ mixing angle (known  \cite{PDG}
to be $\theta_{\omega\phi}=36^{\circ}$) of traditional
$SU(3)$-based scheme. In other words, the $q$-deformation of flavor
symmetries {\em supersedes} the issue of singlet mixing.

For $3<n\le 6$ the scheme works as well.
 Again, only masses of singlets ${\omega}_{15}$,
${\omega}_{24}$, ${\omega}_{35}$ from $(n^2-1)$-plets
of $U_q(u_n)$ involve $q$-dependence. As result, we get
the $q$-analog (with isodoublet $D^*$)
\begin{eqnarray}
 {}  &  \frac{[4]}{[3]} m_{\omega_{15}} {+} \left(
2\frac{[4]^2}{[3]^2}-\frac{8}{[2]}\frac{[4]}{[3]}-
                           \frac{[4]}{[3]}+4 \right)\ m_{\rho} =
                                         \nonumber   \\
{}  &  = 2\ m_{D^*} {+}
 \left( 2\frac{[4]^2}{[3]^2}-\frac{8}{[2]}\frac{[4]}{[3]} + 2 \right)
                    m_{K^*} ,  \label{(4)}
\end{eqnarray}
and $q$-analogs for $n{=}5,6$ (see  \cite{ga-ser,ga-bre,ga-uzh}). Fixing
$q$ by setting $[4]_q{=}[3]_q$ (and $[n]_q{=}[n{-}1]_q$, $n{=}5,6$)
oversimplifies the relations and yields higher analogs of
Okubo's nonet sum rule (isodoublets in r.h.s):
\begin{eqnarray}
\vspace{-3.2mm}
& \hspace{-28mm}  m_{{\omega}_{15}}{+}(5 - 8/[2]_{q_4}) m_{\rho} =
                                                      \nonumber \\
 & \hspace{-8mm}{=}~2\ m_{D^*}+(4 - 8/[2]_{q_4}) m_{K^*} ,  \\ \label{(5)}
& \hspace{-28mm} m_{{\omega}_{24}}{+}(9 - 16/[2]_{q_5}) m_{\rho} =
                                                  \nonumber \\
 &{=}~2\ m_{{D_b}^*}+(4-8/[2]_{q_5})(m_{D^*}+m_{K^*}) ,
                                                            \\ \label{(6)}
    & \hspace{-28mm} m_{{\omega}_{35}}{+}(13-24/[2]_{q_6}) m_{\rho} =
                                                   \nonumber \\
 &{=}~2\ m_{{D_t}^*}{+}(4{-}8/[2]_{q_6})
                        (m_{{D_b}^*}{+}m_{D^*}{+}m_{K^*}).   \label{(7)}
\vspace{0.6mm}
\end{eqnarray}
Here the values $q_n=e^{i\pi/(2n-1)}$
(for which $[2]_{q_n}=2\cos\frac{\pi}{2n{-}1}$)
solve eqs. $\ [n]_q-[n\!-\!1]_q=0$.
Like in the case with $m_{{\omega}_8}\equiv m_{\phi},$ it is meant
in (5)-(7) that $J/\psi $ is put in place of ${\omega }_{15},$
$\Upsilon$ in place of ${\omega }_{24},$ toponium in place of
${\omega }_{35}$ (i.e., \underline{no mixing}!).
With experimental masses, eq.(5) holds to within $ 2.6 \%$
and eq.(6) holds with precision $\simeq 0.7 \%$ .

The $q$-polynomials $[n]_q-[n\!-\!1]_q$ have a topological meaning.

\begin{center}
{\it Labelling flavors by knots invariants }
\end{center}

Polynomials $[n]_q\!-\![n\!-\!1]_q\equiv P_n(q),$ by their roots,
reduce $q$-analogs (2), (4) and those for $n{=}5,6$ to realistic
mass sum rules (MSR) (3), (5)-(7). And, due to the property
$(i)\ P_n(q)=P_n(q^{-1}),\ (ii)\ P_n(1)=1,$ they coincide
              \cite{ga-ser,ga-uzh}
with such knot invariants as Alexander polynomials
$\Delta_q\{(2n\!-\!1)_1\}$ of $(2n-1)_1$-torus knots. E.g.,
$[3]_q-[2]_q=q^2+q^{-2}-q-q^{-1}+1\equiv \Delta_q\{ 5_1\} ,$
$[4]_q-[3]_q=q^3+q^{-3}-q^2-q^{-2}+q+q^{-1}-1\equiv \Delta_q\{ 7_1\}$
correspond to the $5_1$- and $7_1$-knots. Since the extra $q$-deuce
in (4) can be linked to the trefoil (or $3_1$-) knot:
$[2]_q -1=q+q^{-1}-1\equiv\Delta_q\{3_1\}$,
{\it all the} $q$-{\it dependence} in masses of $\omega_{n^2-1}$,
in coefficients of (2),(4) and of higher $q$-analogs,
is expressible through Alexander polynomials. Namely,
$
\frac{[3]_q}{[2]_q}=1+\frac{\Delta \{5_1\}}{[2]_q}=
1+\frac{\Delta \{5_1\}}{\Delta \{3_1\} +1},  
$
$
\frac{[4]_q}{[3]_q}=1+\frac{\Delta \{7_1\}}{[3]_q}=
1+\frac{\Delta \{7_1\}}{\Delta \{5_1\}+\Delta \{3_1\} +1},
$
etc.
The values $q_n$ are thus roots of respective Alexander polynomials.
For each $n$, just the {'senior'} (numerator) polynomial
in $\frac{[3]_q}{[2]_q}$, $\frac{[4]_q}{[3]_q}$ and
$\frac{[n]_q}{[n{-}1]_q}$, $n=5,6$,  serves to 'single out',
by its root, the corresponding MSR from $q$-deformed analog.

Thus, the $q$-parameter for each $n$ is fixed {\it rigidly}
as a root $q_n$ of $\Delta \{ (2n-1)_1\},$ contrary to choice of $q$
by fitting in other phenomenological applications        \cite{iwao}.
Here, using flavor $q$-algebras along with 'dynamical' $q$-algebras
according to $U_q(u_{n})\subset U_q(u_{n+1})$, we gain: the torus knots
$5_1,\ 7_1,\ 9_1,\ 11_1$ \ are put into correspondence \cite{ga-bre,ga-uzh}
with vector quarkonia $s\bar s$, $c\bar c$, $b\bar b$, and $t\bar t$
respectively.  The polynomial $P_{n}(q)\equiv [n]_q-[n\!-\!1]_q$ by its
root $q_n=q(n)$ \underline{determines the value} of $q$-parameter for
each $n$ and thus serves as {\it defining polynomial} for the
MSR/quarkonium/flavor corresponding to $n$.  Hence, the use of
$q$-algebras suggests a possibility of
{\it topological labeling of flavors}:\ fixed number $n$ corresponds
to $2n\!-\!1$ overcrossings of 2-strand braids whose closure gives
these $(2n-1)_1$-torus knots. With the form $(2n\!-\!1,2)$ of same
torus knots this means the correspondence
$n\leftrightarrow w\equiv 2n{-}1,$ $w$ being the winding number
around one of the two basic cycles  on torus.

\section{Octet baryon mass sum rules:
         best candidates from $q$-deformation}
Using $U_q(su_n)$, the $q$-deformed mass relation
\begin{eqnarray}
\vspace{-0.6mm}
   {} & [2]M_N+\frac{[2] M_{\Xi}}{[2]-1} = [3] M_{\Lambda }
+ \Bigl ( \frac{[2]^2}{[2]-1}-[3]  \Bigr )
          M_{\Sigma }                            \nonumber \\
  {} &  +\frac{A_q}{B_q}\left( M_{\Xi } + [2] M_N -
   [2]M_{\Sigma } - M_{\Lambda } \right)             \label{(8)}
 \end{eqnarray}
was obtained  \cite{ga-bre,ga-uzh} where $A_q$, $B_q$ are certain
polynomials of $[2]_q$ with non-overlapping sets of zeros.
This $q$-analog  yields, as three particular cases, the familiar
Gell-Mann - Okubo mass relation (in the 'classical' case of $q=1$)
and two new MSRs of improved accuracy    \cite{ga-bre,ga-uzh,ga-io}:
\vspace{0.6mm}
\begin{eqnarray}
 & \hspace{-9.8mm}   M_N+M_\Xi=\frac32 M_\Lambda+
\frac12M_\Sigma ,  \hspace{10mm} (0.58\%)  \hspace{-4mm}
                                                       \label{(9)} \\
 & \hspace{-1.1mm} M_N+\frac{1\!+\!\sqrt{3}}{2}M_\Xi=
                    \frac{2 M_\Lambda}{\sqrt{3}}+
\frac{9\!-\!\sqrt{3}}{6}M_\Sigma ,    \hspace{1mm} (0.22\%) \ \ \
                                                        \label{(10)} \\
 & \hspace{-6.7mm} M_N+\frac{M_{\Xi}}{[2]_{q_7}\!-\!1}=
\frac{M_{\Lambda}}{[2]_{q_7}\!-\!1}+M_{\Sigma} .
         \hspace{7mm}  (0.07\%)  \hspace{-0.5mm}        \label{(11)}
\end{eqnarray}
\vspace{0.6mm}
Different dynamical representations, after calculation,
produce in (8) differing pairs $A_q$, $B_q$.
Each $A_q$ contains the factor $([2]_q-2)$ i.e.,
the 'classical' zero $q=1$, and some other nontrivial zeros.
Eqs. (10), (11) result from two different dynamical representations
$D^{(1)}$ resp. $D^{(2)}$ producing $A_q^{(1)}$ resp. $A_q^{(2)}$
which possess zeros $q_6=e^{{\rm i}\pi/6}$ resp. $q_7=e^{{\rm i}\pi/7}$.
The choise (11), i.e. $q_7=e^{{\rm i}\pi/7}$, provides the best
mass sum rule.\footnote{
The value $q_7$ is linked \cite{ga-uzh,ARW} to the Cabibbo angle:
$\frac1i\ln{q_7}{\equiv}\theta_{\bf 8}=
\frac{\pi}{7}{=}2\theta_{\scr{C}}$ (see also sec.~7 below).}

Sum rule (10) was first derived \cite{ga-bre} from a dynamical
representation (irrep) $D^{(1)}$ of $U_q(u_{4,1})$.
However, the 'compact' dynamical $U_q(u_5)$ is equally well suited.
Among the admissible dynamical irreps there exist an entire series
of irreps (numbered by integer $m$, $6\le m<\infty $) which produce
infinite set of MSRs, each given by the first line in (8)
with $q_m$ put for $q$, where $q_m=e^{{\rm i}\pi/m}$ guarantees
vanishing of $\frac{A_q}{B_q}$.
{\it Each of these MSRs shows better agreement with data} than the
classical GMO one.  To illustrate, few cases from the infinite set
are shown in the table, the 1st row of which being the classical GMO
with $q_{\infty}=1$.
\vspace{1.0mm}
\begin{center}
\begin{tabular}{ccc}
\hline
$\theta=\frac{\pi}{m}$  &  (RHS$-$LHS),\  $MeV$  &
       $\frac{\left| {\rm RHS}-{\rm LHS}\right|}{{\rm RHS} }, \%$ \\
\hline
$\pi /\infty$ &  26.2    &   0.58 \\
$\pi /30$    &   25.42   &   0.56 \\
$\pi /12$    &   20.2    &   0.44 \\
$\pi /8$     &   10.39   &   0.23 \\
$\pi /7$     &    3.26   &   0.07  \\
$\pi /6$     &  -10.47   &   0.22  \\
\hline
\end{tabular}
\end{center}
\vspace{1.0mm}
We thus gain that a 'discrete choice' becomes possible
instead of usual fitting; the $q$-polynomials $A_q$
due to zeros $q_m$ serve as {\it defining} polynomials
for the corresponding MSRs.

\begin{center}
{\it Quark mass ratio in terms of baryon masses }
\end{center}

Since $[2]_{q_7}=2\cos\frac\pi7$, the MSR (11)
takes the equivalent form of ``modified average''
\begin{equation}
\frac{ M_{\Xi}-M_{N}+M_{\Sigma}-M_{\Lambda} }{ 2\cos(\pi/7) } =
            M_{\Sigma}-M_{N} .                          \label{(12)}
\end{equation}
From (12) with $\frac{\pi}{7}=2\theta_{\scr C}$ (see footnote 2),
using the famous relation      \cite{cab-ma}
$\tan^2\theta_{\rm C}=m_d/m_s$ we infer a new formula giving quark
mass ratio in terms of (very precisely known) octet baryon masses:
\[
\frac{m_s}{m_d} = \frac{ 3 M_{\Sigma}-M_{\Lambda}-3M_{N}+M_{\Xi} }
                       { M_{\Sigma}+M_{\Lambda}-M_{N}-M_{\Xi} }
                   = 18.63\pm 0.16 .
\]
Numerically, the obtained ratio is in nice agreement with
the value $\frac{m_s}{m_d}{=}18.9\pm 0.8$ given in \cite{leu00}.

\section{Mass sum rules for decuplet baryons:
         the $q$-analog matches empirical masses}
In the case of $SU(3)$-decuplet baryons ${\frac32}^+$,
1st order symmetry breaking yields         \cite{8-fold}
equal spacing rule (ESR) for isoplet members
in ${\bf 10}$-plet. Empirically,
$M_{\Sigma^*}{-}M_{\Delta},$ $ M_{\Xi^*}{-}M_{\Sigma^*}$ and
$M_{\Omega}{-}M_{\Xi^*}$ show sensible deviation from ESR:
$152.6~MeV\leftrightarrow 148.8~MeV\leftrightarrow 139.0~MeV$.
The other mass relation known long ago     \cite{oku-ko},
\begin{equation}
(M_{\Sigma^{*}}-M_{\Delta}+M_{\Omega}-M_{\Xi^{*}})/2
= M_{\Xi^{*}}-M_{\Sigma^{*}} ,                            \label{(13)}
\end{equation}
accounts 1st and 2nd order of $SU(3)$-breaking and holds
only slightly better than the ESR.

On the contrary, use of the $q$-algebras $U_q(su_n)$ instead of
$SU(n)$ leads to sizable improvement. From evaluations of
decuplet masses in two particular irreps of the dynamical
algebra $U_q(u_{4,1})$, the $q$-deformed mass relation
\begin{equation}
\frac{ M_{\Sigma^*}-M_{\Delta}+M_{\Omega}-M_{\Xi^*} }
     { 2\cos\theta_{\bf 10} }
= M_{\Xi^*}-M_{\Sigma^*}
                                                       \label{(14)}
\end{equation}
was derived   \cite{GKT}. As proven there, this mass relation is
\underline{universal} - it results from any admissible irrep (which
contains $U_q(su_3)$-decuplet embedded in {\em 20}-plet of $U_q(su_4)$)
of the dynamical $U_q(u_{4,1})$.  With empirical masses   \cite{PDG},
the formula (14) holds remarkably for
$\theta_{\bf 10}\simeq\frac{\pi }{14}$
(in fact, $\theta_{\bf 10}=\theta_{\scr{C}}$,
see footnote 2 and sec. 7 below).

The universality of $q$-analog (14) extends also to all
admissible irreps of the {\it 'compact'} dynamical $U_q(su_5)$.
Say, within a dynamical irrep $\{ 4 0 0 0 \}$ of $U_q(su_5)$
calculation yields:
$M_{\Delta}=M_{\bf 10} +  \beta ,\ $
$M_{\Sigma^*}=M_{\bf 10} + [2] \beta +  \alpha ,\ $
$M_{\Xi^*}=M_{\bf 10} + [3] \beta + [2] \alpha ,\ $
$M_{\Omega}=M_{\bf 10} + [4] \beta + [3] \alpha ,\ $
from which (14) stems.  With hypercharge $Y$, all
four masses are comprised by single formula for
$M_{D_i} \equiv M\bigl( Y(D_i) \bigr )$:
\begin{equation}
M_{D_i}{=}M_{\bf 10} {+} {\alpha} [1\!-\!Y(D_i)]_q
             + {\beta} [2\!-\!Y(D_i)]_q ,               \label{(15)}
\end{equation}
with explicit dependence on $Y$.
If $q=1$, this reduces to $M_{D_i} = \tilde{M}_{\bf 10} + a~Y\!(D_i)$,
i.e. {\it linear dependence on hypercharge}
(or strangeness) where $ a = - \alpha - \beta ,\ $
$\tilde{M}_{\bf 10} = M_{\bf 10} + \alpha + 2 \beta.$

\section{Highly nonlinear $SU(3)$-breaking effects in baryon masses}
Formula (15) involves {\it highly nonlinear dependence} of mass on
hypercharge (for decuplet, $Y$ alone causes $SU(3)$-breaking).
Since for the $q$-number $[N]_q$ we have
$[N]_q=q^{N-1}+q^{N-3}+\ldots +q^{-N+3}+q^{-N+1}$ ($N$ terms),
this shows exponential $Y$-dependence of masses. Such
high nonlinearity makes (14) and (15) radically different from
the result (13) of traditional treatment accounting linear and
quadratic effects in $Y$.

For octet baryon masses, {\em nonpolynomiality}
in $SU(3)$-breaking effectively accounted by the model was
explicitly shown in     \cite{ga-io}.
For this, one analyses the expressions for isoplet
masses with explicit dependence on hypercharge $Y$ and isospin $I$,
through $I(I+1)$.  Typical matrix element contributing to
octet baryon masses contains terms such as
$\left({[Y/2]_q[Y/2\!+\!1]_q-[I]_q[I\!+\!1]_q}\right)$  or
$\left({[Y/2-1]_q[Y/2-2]_q-[I]_q[I+1]_q}\right)$  (with multipliers
depending on irrep labels $m_{\scr{15}},m_{\scr{55}}$), which show
explicitly the dependence on hypercharge and the factor
$[I]_q[I+1]_q$ $q$-deforming the $SU(2)$ Casimir.  From definition
of $q$-bracket $[n]=\frac{\sin(nh)}{\sin(h)}$, $ q\!=\!\exp({\rm i}h)$,
it is clearly seen that baryon masses depend on hypercharge $Y$ and
isospin $I$ (hence, on $SU(3)$-breaking effects) in
highly nonlinear - {\it nonpolynomial} - fashion.

The ability to account highly nonlinear
$SU(3)$-breaking effects by applying
quantum analogs $U_q(su_n)$ of usual flavor symmetries
is much alike the fact shown in    \cite{lor-we} that,
by exploiting appropriate {\em free} $q$-deformed
structure one is able to efficiently study the properties of
(undeformed) quantum-mechanical systems with complicated
interactions.

\section{Using the Hopf-algebra structure}
As demonstrated, our approach supplies a plenty of $q$-analogs
(with different pairs $A_q,B_q$) of the form (8).
A completely different, as regards (8), version of
$q$-deformed analog can be derived    \cite{ga-io}
using for the symmetry breaking term in mass operator a component
of $q$-{\em tensor operator}. This implies usage of the Hopf algebra
structure (comultiplication, antipode) of the quantum algebras
$U_q(su_n)$, through the $q$-tensor operators $(V_1,V_2,V_3)$ resp.
$(V_{\bar{1}},V_{\bar{2}},V_{\bar{3}})$ formed from elements of
$U_q(su_4)$ and transforming as ${{\bf 3}}$ resp. ${{\bf 3}^*}$
under the adjoint action of $U_q(su_3)$. With the Cartan elements
$H_1, H_2$, denoting $[X,Y]_q {\equiv} XY{-}qYX$, the components
$(V_1,V_2,V_3)$ and $(V_{\bar{1}},V_{\bar{2}},V_{\bar{3}})$
can be found explicitly          \cite{ga-io},   e.g.,
 $V_2=[E_2^+,E_3^+]_q q^{H_1/6-H_2/6}$,  $V_3=E_3^+ q^{H_1/6+H_2/3},$
and  $V_{\bar{3}}=q^{H_1/6+H_2/3} E_3^-$.
 The mass operator is given as
\[
\begin{array}{ll}
{} & \hat M = \hat M_0+\hat M_8
= M_0 {\bf 1} + \alpha V_8^{(1)} + \beta V_8^{(2)}   \\
{} & \hspace{5mm}
= M_0 {\bf 1} +\alpha V_3 V_{\bar{3}} + \beta V_{\bar{3}} V_3
\end{array}
\]
where $\hat{M}_0$ is $U_q(su_3)$-invariant and $\hat{M}_8$ transforms
as $I\!=\!0,Y\!=\!0$ component of tensor operator of ${\bf 8}$-irrep
of $U_q(su_3)$, and it is taken into account that the irrep ${\bf 8}$
occurs twice in the decomposition of ${{\bf 8}}\otimes{{\bf 8}}$.
Besides, the isosinglet operators $V_{3}V_{\bar{3}}$ and
$V_{\bar{3}}V_{3}$ arise in accordance with
${{\bf 3}}\otimes {{\bf 3}}^*\!=\!{{\bf 1}}\oplus {{\bf 8}},$
${{\bf 3}}^*\otimes {{\bf 3}}\!=\!{{\bf 1}}\oplus {{\bf 8}}$.

The final form of mass operator, with redefined $M_0,\alpha,\beta$, is
\begin{equation}
\hat M = M_0 {\bf 1}
 +\alpha E_3^+ E_3^- q^Y + \beta E_3^- E_3^+ q^Y               \label{16}
\end{equation}
where the hypercharge $Y=(H_1+2H_2)/3$.
Matrix elements with $\hat M$ from (16) are evaluated by
embedding ${\bf 8}$ in a particular irrep of $U_q(su_4).$
Evaluation of baryon masses, say,  within the irrep ${\bf 15}$
of $U_q(su_4)$ yields:
$ M_N=M_0+\beta q ,\ M_\Sigma= M_0 ,\  M_\Lambda =M_0 +
  \frac{[2]}{[3]} (\alpha+\beta) ,\ M_\Xi=M_0+\alpha q^{-1}$.
Excluding $M_0,\alpha$, $\beta$, one finds:
\begin{equation}
[3] M_\Lambda + M_\Sigma=[2] (q^{-1} M_N+q M_\Xi) .            \label{17}
\end{equation}
The $q$-parameter now {\it can be fixed by a fitting only} and,
for each of values $q_{1,2}=\pm 1.035$, $q_{3,4}=\pm 0.903 \sqrt{-1}$,
the $q$-MR (17) indeed holds within experimental uncertainty.

\section{The link: $q$-parameter $\leftrightarrow$ Cabibbo angle}
For pseudoscalar (PS) mesons, the generalization   \cite{oku70}
of GMO-formula
\begin{equation}
f_\pi^2 m_\pi^2 + 3 f_\eta^2 m_\eta^2 = 4 f_K^2 m_K^2      \label{18}
\end{equation}
involves decay constants as coefficients.
On imposing the constraint\footnote{  It leads to the single
dimensionless quantity $\frac{f_K}{f_\pi}$ involved in
the multipliers of masses.}
${f^{2}_\pi} + {3} {f^{2}_\eta} = {4}{f^{2}_K}$ it becomes
\begin{equation}
m_\pi^2 +      
\Bigl( 4 \frac{f_K^2}{f_\pi^2} - 1 \Bigr) m_\eta^2
= 4 \frac{f_K^2}{f_\pi^2} m_K^2 ,                          \label{19}
\end{equation}
to be compared with our $q$-{\it analog} (2) rewritten for PS mesons
(with masses squared):
\begin{equation}
m_\pi^2 + \frac{[3]}{2 [2] - [3]} m_{\eta_8}^2
= \frac{2 [2]}{2 [2] - [3]} m_K^2 .                        \label{20}
\end{equation}
This holds for (the mass of) {\it physical} $\eta$-meson put
for $\eta_8$ (i.e., no mixing), {\it at properly fixed} $q=q_{\scr{PS}}$.

The two extensions (19) resp. (20) both reduce to standard GMO
in the corresponding limits $\frac{f_K}{f_\pi}\to 1$ resp. $q\to 1$.
From the identification
\begin{equation}
{f_K^2}/{f_\pi^2} \
      \leftrightarrow \  {\frac12 [2]}/({2[2]\!-\![3]}),    \label{(21)}
\end{equation}
using $[3]_q=[2]_q^2-1$ and the notation
${\xi}_{\pi,K}\equiv({4 f^2_K/f^2_\pi})^{-1}$, we get
\begin{equation}
[2]_{\pm}= 1 -
{\xi}_{\pi,K}\pm\sqrt{\bigl(1-{\xi}_{\pi,K}\bigr)^2 + 1}.    \label{(22)}
\end{equation}
The ratio $f_K/f_\pi$ is expressible through the Cabibbo angle, e.g.,
by the formula   $\tan^2\theta_{\scr{C}} = \frac{m_\pi^2}{m_K^2}
\Bigl[\frac{f_K}{f_\pi}-\frac{m_\pi^2}{m_K^2}\Bigr]^{-1}$
       (see     \cite{oakes}).
 With (21), (22) this implies: the {\it deformation parameter}
$q_{\scr{PS}}$  {\it is directly connected with the Cabibbo angle}.

One can arrive at similar conclusion in another way. In    \cite{isa-po},
the $q$-deformed lagrangian for gauge fields of the Weinberg - Salam (WS)
model, invariant against quantum-group valued gauge transformations,
was constructed. The formula
$F^0_{\mu\nu}\ =\ {\rm Tr}_q(F_{\mu\nu})\ [2(q^2 + q^{-2})]^{-1/2}
 = \ B_{\mu\nu} \cos\theta + F^3_{\mu\nu}\sin\theta $  obtained therein,
along with expression for $F^3_{\mu\nu}$ and the relation
\begin{equation}
    \tan\theta = (1-q^2)/(1+q^2) ,                          \label{23}
\end{equation}
exhibits mixing of the $U(1)$-component $B_\mu$ with third (nonabelian)
component $A^3_\mu$.  Forming new potentials
$\tilde{A}_\mu = B_\mu\cos\theta + A^3_\mu \sin\theta,$
$\ Z_\mu=-B_\mu \sin\theta + A^3_\mu \cos\theta$  yields
physical photon $\tilde{A}_\mu$ and $Z$-boson of WS model,
where $\theta=\theta_{\rm\scr{W}}$, i.e., the Weinberg angle
(at $\theta = 0$ the potentials $B_\mu$ and $A^3_\mu$ get
completely unmixed whereas nonzero $\theta$, i.e., nontrivial
$q$-deformation provides proper mixing inherent for the WS model).
To summarize: {\em weak mixing is adequately modelled by
the $q$-deformation}. That is, the $q$-deformation is able to
realize proper weak mixing of gauge fields and provides explicit
connection of the weak angle and the deformation parameter $q$,
see eq.(23).

On the other hand, the relation found in   \cite{palle}
\begin{equation}
\theta_{\scr{W}}
         = 2 (\theta_{12} + \theta_{23} + \theta_{13})      \label{24}
\end{equation}
connects $\theta_{\scr{W}}$ with the Cabibbo angle
$\theta_{12}\equiv \theta_{\scr{C}}$ (and the angles
$\theta_{13}, \theta_{23}$ of mixing with 3rd family,
that will be neglected).
The eqn. (24) is important as it links apparently different
mixings: in the {\it bosonic} (interaction) versus
{\it fermionic} (matter) sectors of the electroweak model.

Combining (23) and (24)
we conclude: the Cabibbo angle can be linked with
$q$-parameter of a quantum-group (or quantum-algebra)
based structure {\it applied in the fermion sector}.
Hence, there must exist a direct connection of the $q$-parameter
in (12), (14) with the fermion mixing angle. Setting
$\theta_{\bf 10}=g(\theta_{\scr{C}})$ and
$\theta_{\bf 8}=h(\theta_{\scr{C}})$ we find for $g(\theta_{\scr{C}})$
and $h(\theta_{\scr{C}})$ the following:
\begin{equation}
\theta_{\bf 10}=\theta_{\scr{C}},
\hspace{18mm}  \theta_{\bf 8}=2~\theta_{\scr{C}}.         \label{(25)}
\end{equation}
With $\theta_{\bf 8}=\frac{\pi}{7}$, see (12),
this suggests for Cabibbo angle the exact value
$\frac{\pi}{14}$.

\begin{center}
{\it Cabibbo mixing from noncommutative extra dimensions?}
\end{center}

Quantum groups and the related quantum algebras provide
necessary tools in constructing covariant differential calculi and
particular noncommutative geometry on quantum spaces   \cite{frt},
e.g, quantum vector spaces, quantum homogeneous spaces.

The direct link found between the Cabibbo angle
$\theta_{\scr{C}}{=}\frac{\pi}{14}$ and the $q$-parameter which
measures strength of $q$-deformation for the $q$-algebras $U_q(su_n)$
used for flavor symmetry, can be viewed \cite{ARW} as indicating
towards noncommutative-geometric origin of fermion mixing. The exact
value $\theta_{\scr{C}}{=}\frac{\pi}{14}$ of the Cabibbo angle would
then serve as the {\it noncommutativity measure} of relevant quantum
space, responsible for the mixing and explicitly as yet unknown,
in extra dimensions whose number should be not less than 2.

\section{ Mass relations from anyonic realization of $U_q(su_N)$ }

Necessary setting adopted from   \cite{ler-sci} includes
lattice angle functions $\theta_{\gamma}({\bf x},{\bf y})$ and
$\theta_{\delta}({\bf x},{\bf y})$ for the two opposite
($\gamma$- and $\delta$- ) types of cuts and the related definition
of ordering of sites on the lattice (${\bf x} > {\bf y}$ or
$ {\bf y} > {\bf x}$). Accordingly, the two types of statistical operator,
$K_i({\bf x}_{\gamma})$ and $K_i({\bf x}_{\delta})$,
are formed using $N$ sorts of lattice fermions
$c_i({\bf x})$, $\ c_i^{\dagger}({\bf x})$, $\ i=1,...,N,$
obeying usual (lattice) anticommutation relations (ARs), as
\begin{equation}
K_j({\bf x}_{\gamma})=
\exp\bigl( {\rm i} \nu
    \sum_{{\bf y}\ne{\bf x}}\theta_{\gamma}({\bf x},{\bf y})
c_j^{\dagger}({\bf y})c_j({\bf y}) \bigr)                 \label{(26)}
\end{equation}
and similarly for $K_i({\bf x}_{\delta})$.
In terms of them, the two types of
anyonic oscillators are given as   \cite{ler-sci}
\[
a_i({\bf x}_{\gamma})= K_i({\bf x}_{\gamma})c_i({\bf x}), \ \ \ \
a_i({\bf x}_{\delta})= K_i({\bf x}_{\delta})c_i({\bf x}).
\]
The relations of permutation (PRs) obtained for anyonic oscillators
include simple ARs, and also nontrivial PRs involving
the deformation parameter $q$ (the latter is connected with the
statistics parameter $\nu$ in eq. (26) as: $\ q=\exp({\rm i}\pi\nu)$).
For instance, the braiding properties are described by the following
nontrivial PRs (${\bf x}\ne {\bf y}$):
\[
\begin{array}{ll}
a_i({\bf x}_{\gamma})a_i({\bf y}_{\gamma})
+q^{-{\rm sgn}({\bf x}-{\bf y})}
a_i({\bf y}_{\gamma})a_i({\bf x}_{\gamma}) =0, & {}         \\
      a_i({\bf x}_{\gamma})a_i^{\dagger}({\bf y}_{\gamma})
       +q^{{\rm sgn}({\bf x}-{\bf y})}
          a_i^{\dagger}({\bf y}_{\gamma})a_i({\bf x}_{\gamma}) =0.
\end{array}
\]
The basic fact proven in  \cite{ler-sci} states that generating
elements $A_{j,j+1}$, $A_{j+1,j}$ and $H_j$ realized bilinearly
in terms of anyonic oscillators $a_i({\bf x}_{\gamma})$,
$a_i^{\dagger}({\bf y}_{\gamma})$ satisfy the defining
relations  \cite{dr-j,frt} of the quantum algebra $U_q(su_N)$.
Similarly, dual realization in terms of  $a_i({\bf x}_{\delta})$,
$a_i^{\dagger}({\bf y}_{\delta})$ does also exist. On this basis,
within anyonic realization of $U_q(su_N)$, one can
explicitly construct    \cite{ga-io2}  both basis vectors
for hadronic irreps and hadron mass operator.
Starting point is the heighest weight vector (HWV) of
the irrep $\{4000\}$ of 'dynamical' $U_q(su_5)$ which is
of the form $|1111\rangle$ in the notation
$|n_1n_2n_3n_4\rangle$ for the state vector, that means
$
a_{n_1}^{\dagger}({\bf x}_{1\gamma})
a_{n_2}^{\dagger}({\bf x}_{2\gamma})
a_{n_3}^{\dagger}({\bf x}_{3\gamma})
a_{n_4}^{\dagger}({\bf x}_{4\gamma})|0\rangle .
$
All basis state vectors of baryons $\frac32^+$ are constructed,
by acting with lowering generators, in accordance with the chain
of $q$-algebras
$U_q(su_3)\subset U_q(su_4)\subset U_q(su_5)$
and respective chain of irreps $[30]\subset [300]\subset \{4000\}$.
For isoquartet baryon $|\Delta^{++}\rangle$ one finds
$ 
\frac{1}{\sqrt{[4]}}
(|5111\rangle + q^{-1}|1511\rangle +
              q^{-2}|1151\rangle + q^{-3}|1115\rangle),
$
and similarly for $|\Sigma^*\rangle$, $|\Xi^*\rangle$,
$|\Omega^-\rangle$. The dual basis
$\widetilde{|\Delta^{++}\rangle}$, etc., obtained by acting on
the HWV with lowering operators in dual anyonic realization,
is also needed. Masses $M_{D_i}$ of baryons $D_i$ are calculated
within the dynamical $U_q(su_5)$-irrep $\{4000\}$ as
$M_{D_i}=\widetilde{\langle D_i|}\hat{M}|D_i\rangle $
(with mass operator formed from anyonic operators) to yield:
$M_{\Delta}=M_{\bf 10}+\beta$,
$M_{\Sigma^*}=M_{\bf 10}+[2]_q\alpha+[2]_q\beta$,
$M_{\Xi^*}=M_{\bf 10}+[2]_q^2\alpha+[3]_q\beta$,
and $M_{\Omega^*}=M_{\bf 10}+[2]_q[3]_q\alpha+[4]_q\beta$. One easily
checks that these masses satisfy the relation (14).
This proves applicability         \cite{ga-io2}  of
quantum algebras and their irreps for treating
hadron mass relations {\it within anyonic realization}.

\section{Algebras of $q$-oscillators, $q$-Bose gas and
         two-pion (two-kaon) correlations }

The model of ideal gas of $q$-bosons based on the algebra of
$q$-deformed oscillators either of Biedenharn-Macfarlane (BM)
type      \cite{BM}  or  Arik-Coon (AC) type      \cite{AC},
was recently utilized within the approach aimed to describe
       \cite{AGI,AGI2}  unusual properties of 2-particle correlations
of identical pions or kaons produced in heavy ion collisions.
The approach yields clear predictions based on explicit expressions
for the intercept $\lambda$ (dependent on temperature, particle mass,
pair mean momentum, {\it and the deformation parameter} $q$).

To obtain needed observables, one evaluates thermal averages
$\langle A \rangle ={\rm Sp}(A\rho )/{\rm Sp}(\rho)$,
$\rho = e^{-\beta H}$, where the Hamiltonian $H={\sum}\omega_i N_i$
and $\beta=1/T$.
With $b^\dagger_i b_i=[N_i]_q$ and $q+q^{-1}=2\cos\theta$,
the $q$-deformed distribution function results for BM-type $q$-bosons as
\vspace{-0.3mm}
\begin{equation}
\langle b_i^\dagger b_i \rangle=\frac{e^{\beta\omega_i}-1}
{e^{2\beta\omega_i}-2\cos\theta~e^{\beta\omega_i}+1} .      \label{27}
\end{equation}
\vspace{-0.3mm}
At $\theta{=}0$ (or $q{=}1$), it yields Bose-Einstein (B-E) distribution,
since $q{=}1$ recovers usual bosonic commutation relations.
As seen, deviation of $q$-distribution (27) from the quantum
B-E distribution tends towards the Maxwell-Boltzmann one.
 This means reducing of quantum statistical effects.
For kaons, whose mass $m_K>3m_{\pi}$, analogous curve gets closer
(than pion's one) to the B-E distribution.
For AC-type $q$-bosons, the $q$-distribution is especially simple:
$
\langle b_i^\dagger b_i \rangle=\frac{1}{e^{\beta\omega_i}-q} .
$

To obtain explicitly the intercept $\lambda$ of two-particle
correlations one starts with the defining ratio
$\lambda + 1 = \langle b^\dagger b^\dagger b b\rangle /
(\langle b^\dagger b \rangle)^2$,  calculates the two-particle
distribution $\langle b^\dagger b^\dagger b b\rangle$ and takes
into account the $\langle b^\dagger b \rangle$. 
The result for AC-type $q$-bosons reads
$\lambda=q-\frac{q(1-q^2)}{e^{\omega/T}-q^2}$, $\ -1\le q\le 1$,
and for BM-type $q$-bosons, with
${\cal F}(\beta\omega)\equiv\cosh(\beta\omega)$, it is
\begin{equation}
\lambda\!=\!-1\!+\!\frac{ 2\cos\theta\
({\cal F}(\beta\omega)\!-\!\cos\theta)^2 }
{({\cal F}(\beta\omega)\!-\!1)({\cal F}(\beta\omega)\!-\!2\cos^2\theta+1)}.
                                                           \label{28}
\end{equation}
Both (\ref{27}), (\ref{28}) are real owing to the sum $q+q^{-1}$.

The intercept $\lambda$, with $\omega=(m^2+{\bf K^2})^{1/2}$,
shows a remarkable feature: asymp\-to\-tically, at large
mean momentum of pion (kaon) pair
and fixed temperature,  $\lambda$ tends to a constant
given by the $q$-parameter: $\lambda^{\rm as}=q\ $ ($q$ real)
for the AC-type $q$-bosons, and
\vspace{-0.6mm}
\begin{equation}
\lambda^{\rm as}=2\cos\theta-1, \ \ \ \ \ \ \ \
                       \theta=\frac1i~{\rm ln}~q,       \label{(29)}
\end{equation}
\vspace{-0.4mm}
for the BM-type $q$-bosons.

As conjectured in \cite{AGI2}, correlations of pions and kaons are
determined by the same value of $q$ (a kind of universality).
Then, experimentally one should observe the tendency of merging
$\lambda(\pi)$ and $\lambda(K)$ at large enough mean momenta, i.e.,
$\lambda^{\rm as}(\pi)=\lambda^{\rm as}(K)$. Preliminary results
of recent RHIC/STAR experiment give three values   \cite{laue}
$\lambda_1(\pi^-)$, $\lambda_2(\pi^-)$ and $\lambda_3(\pi^-)$ for
the $\pi^-$-intercept, obtained by averaging over three intervals of
transverse momenta (in MeV/c):  $\ (125 \div 225)$, $\ (225 \div 325)$,
$\ (325 \div 450)$, and by integrating over rapidity in the range
$-0.5 \le y \le 0.5$.

  \begin{figure}[t]
\begin{center}
\vspace{-0.01cm}
\hspace{-2mm}   {
\epsfig{file=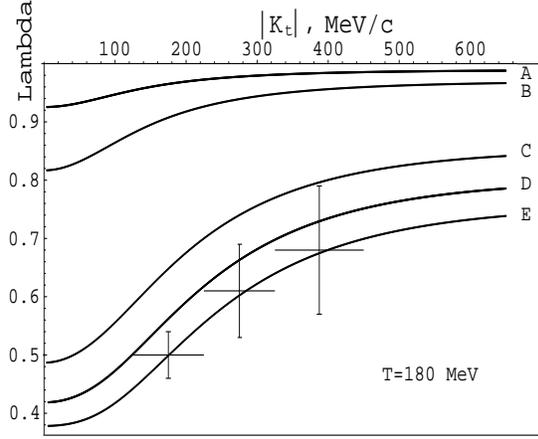,height=5cm,width=6.8cm,angle=0}      }
\vspace{-0.6cm}
\caption  { The transverse momentum $|{\bf K}_t|$ dependence of the
intercept $\lambda $ of two-pion correlation at fixed $T=180$ MeV
and fixed deformation parameter $q=\exp(i\theta)$ :
 \  A) $\theta = 6^\circ$,  \ B) $\theta = 10^\circ$,
 \  C) $\theta = 22^\circ$,
 \  D) $\theta =  25.7^\circ$ (i.e., $2 \theta_{\rm C}$),
 \  E) $\theta = 28.5^\circ$ .}
\vspace{-1.5cm}
\end{center}
\end{figure}

In Fig.1 the three values $\lambda_j(\pi^-)$, $j{=}1,2,3$,
with error bars, are shown along with five curves
for the intercept $\lambda$ which correspond to fixing in (28)
different values of the deformation angle $\theta$,
all curves being at the temperature $T=180$ MeV.  One can see
remarkable agreement with data of the curve E obtained at
$\theta=28.5^{\circ}$. The other interesting curve D corresponds
to $\theta=\pi/7\simeq 25.7^{\circ}$ (twice the Cabibbo angle,
see footnote 2 and eq.(25)). At the same temperature $T=180$ MeV,
the curve D agrees (within error bar) with the points $\lambda_2(\pi)$
and $\lambda_3(\pi)$. However, suffice it to take slightly higher
effective temperature $T\simeq 198$ MeV, and the resulting curve
marked by $\theta\!=\!\pi/7\!=\!2\theta_{\scr{C}}$
{\it respects all the three error bars}. Among different
mixing angles known for hadrons, see \cite{PDG}, only the angle
$2\theta_{\scr{C}}$ seems to be relevant to the discussed topic of
intercept $\lambda(\pi)$. It is tempting to suggest that
just this angle $2\theta_{\scr{C}}$ can be the benchmark of
assumed universality (to be) seen in 2-particle correlations
since, then, $\lambda^{\rm as}(\pi)\vert_{\theta=\pi/7}=
\lambda^{\rm as}(K)\vert_{\theta=\pi/7}=
2\cos\frac{\pi}{7}{-}1\simeq0.80194$.
Insisting on the asymptotical coincidence
$\lambda^{\rm as}(\pi)=\lambda^{\rm as}(K)$
we may predict for kaon intercepts:
{\it at any transverse momentum}, {\it the intercept} $\lambda(K)$
{\it of 2-kaon correlations should not exceed} $0.80194$.

\section{Outlook}

A question naturally arises: does there exist
more intimate connection  between the two
discussed applications - of the quantum algebras $U_q(su_n)$
taken as flavor symmetries, on one hand, and of the algebras
of $q$-deformed oscillators corresponding to discretized momenta
of (correlated) pairs of pions or kaons as produced in
relativistic heavy ion collisions, on the other hand?
The value of $q$-parameter (given by $2 \theta_{\scr C}$)
shared by the two applications in case of octet hadrons
gives a guess for possible physical reason for such a connection
(recall also the well-known fact that generating elements of
$U_q(su_n)$ admit realization in terms of $q$-deformed
oscillators  \cite{BM}).
  Future research possibly involving noncomutative geometry in extra
dimensions should give ultimate answer.

\vspace{1.2mm}
{\bf Acknowledgements.} I express my gratitude to
J.~Lukierski, A.~Isaev for valuable discussion of the results,
to D.~Anchishkin for collaboration concerning the topics in sec.~9.
My sincere thanks are to D.~Sorokin for the invitation
to participate in this nice and useful conference.



\begin{thebibliography}{9}

\bibitem{dr-j} V.~Drinfeld, Sov. Math. Dokl. {\bf 32} (1985) 254.
         M. Jimbo, Lett. Math. Phys. {\bf 10} (1985) 63.

\bibitem{frt}
L.D.~Faddeev, N.~Reshetikhin and L.~Takhtajan, Leningrad Math. J.
{\bf 1} (1990) 193.

\bibitem{ga-ser} A.M.~Gavrilik, J. Phys. {\bf A 27} (1994) L91.
A.M.~Gavrilik and A.V.~Tertychnyj, {\sl preprint} ITP-93-19E, Kiev, 1993.

\bibitem{cabi}
N.~Cabibbo, Phys. Rev. Lett.  {\bf 10} (1963) 531.

\bibitem{ga-bre} A.M.~Gavrilik,
in {\sl "Symmetries in Science VIII"} (Proc. Int. Conf., B.Gruber ed.)
 Plenum, N.Y., 1995, p. 109.\
A.M.~Gavrilik, I.I.~Kachurik, and A.~Tertychnyj,
  {\sl Kiev preprint} ITP-94-34E, 1994, {\tt hep-ph/9504233}.

\bibitem{ga-uzh} A.M.~Gavrilik,
in {\sl "Non-Euclidean Geometry in Modern Physics"}
(Proc. Int. Conf.), Kiev, 1997, p. 183, {\tt hep-ph/9712411}.

\bibitem{8-fold} M.~Gell-Mann and Y.~Ne'eman,
  {\sl The Eightfold Way}, Benjamin, N. Y., 1964.

\bibitem{oku-fi} S.~Okubo, Phys. Lett. {\bf 5} (1963) 165.

\bibitem{PDG} Particle Data Group: Caso C. {\em et al.},
The Europ. Phys. Journ. {\bf C 3} (1998) 1.

\bibitem{iwao} S.~Iwao, Progr. Theor. Phys. {\bf 83} (1990) 363.\
    D.~Bonatsos et al., Phys. Lett. {\bf 251B} (1990) 477.

\bibitem{ga-io} A.M.~Gavrilik and N.Z.~Iorgov, Ukr. J. Phys.
      {\bf 43} (1998) 1526,\ {\tt hep-ph/9807559}.

\bibitem{ARW} A.M.~Gavrilik, in ``Noncommutative Structures in
Mathematics and Physics'', Proc. NATO ARW (ed. by S.~Duplij and J.~Wess),
Kluwer, 2001, p.344,   {\tt hep-ph/0011057}.

\bibitem{cab-ma}
N.~Cabibbo, L.~Maiani, Phys. Lett. {\bf 28} (1968) 131.
R.~Gatto, G.~Sartori and M.~Tonin, Phys. Lett. {\bf 28} (1968) 128.

\bibitem{leu00}
 H.~Leutwyler, {\tt hep-ph/0011049}.

\bibitem{oku-ko}
 S.~Okubo, Phys. Lett. 4 (1963) 14.
 I.~Kokke\-dee,~{\sl The~Quark~Model},~Benjamin,~N.Y.,~1968.

\bibitem{GKT} A.M.~Gavrilik, I.I.~Kachurik, and A.V.~Tertychnyj,
       Ukr. J. Phys.  {\bf 40} (1995) 645.

\bibitem{lor-we} A.~Lorek and J.~Wess,
        Z. Phys. {\bf C 67} (1995) 671,\ {\tt q-alg/9502007}.

\bibitem{oku70} S.~Okubo, in {\sl "Symmetries and quark models"}
(Proc. Int. Conf., ed. by R.Chand), Gordon and Breach, N.Y., 1970.

\bibitem{oakes} R.J.~Oakes, Phys. Lett. {\bf 29} (1969) 683.

\bibitem{isa-po} A.P.~Isaev and Z.~Popowicz, Phys. Lett.
       {\bf B 281} (1992) 271.

\bibitem{palle} D.~Palle, Nuovo Cim. {\bf 109 A} (1996) 1535,
        {\tt hep-ph/9706266}.

\bibitem{ler-sci} A.~Lerda and S.~Sciuto,
         Nucl. Phys. {\bf B 401} (1993) 613.
M.~Frau, M.A.~R.-Monteiro and S.~Sciuto, J. Phys. {\bf A 27} (1994) 801.

\bibitem{ga-io2} A.M.~Gavrilik and N.Z.~Iorgov,
        Ukr. J. Phys. {\bf 45} (2000) 789,\ {\tt hep-ph/9912222}.

\bibitem{BM}
A.J.~Macfarlane,  J.\ Phys.\ {\bf A\ 22}, 4581 (1989).
L.~Biedenharn, J.\ Phys.\ {\bf A\ 22} L873 (1989).

\bibitem{AC}
M.~Arik and D.D.~Coon, J.\ Math.\ Phys.\ {\bf 17} (1976) 524.
D.~Fairlie and C.~Zachos, Phys.\ Lett.\ {\bf 256B} (1991) 43.
S.~Meljanac and A.~Perica,
Mod.\ Phys.\ Lett.\ {\bf A 9} (1994) 3293.

\bibitem{AGI}
 D.V.~Anchishkin, A.M.~Gavrilik and N.Z~Iorgov,
 The Europ. Phys. Journ. A {\bf 7} 229 (2000); {\tt nucl-th/9906034}.

\bibitem{AGI2}
 D.V.~Anchishkin, A.M.~Gavrilik and N.Z~Iorgov,
  Mod.~Phys.~Lett. A{\bf 15} 1637 (2000); {\tt hep-ph/0010019}.

\bibitem{laue} F.~Laue, "Two-particle interferometry at RHIC", 
talk at Quark Matter'2001, to be published in Nucl. Phys. A. 

\end{thebibliography}
\end{document}